\documentclass[twocolumn,showpacs,preprintnumbers,amsmath,amssymb]{revtex4}

\usepackage{graphicx}
\usepackage{dcolumn}
\usepackage{bm}
\usepackage{color}

\definecolor{gold}{rgb}{0.85,.66,0}

\begin{document}

\title{Exciton spin dynamics of colloidal CdTe nanocrystals in magnetic field}

\author{Feng Liu,$^{1}$  A.~V.~Rodina,$^{2}$ D.~ R.~Yakovlev,$^{1,2}$, A.~Greilich,$^{1}$
 A.~A.~ Golovatenko,$^{2,3}$ A.~S.~Susha,$^{4}$   A.~L.~Rogach,$^{4}$  Yu.~G.~Kusrayev,$^{2}$ and M.~Bayer$^{1,2}$}

\affiliation{$^{1}$Experimentelle Physik 2, Technische Universit\"at Dortmund, 44221 Dortmund, Germany}
\affiliation{$^{2}$Ioffe Physical-Technical Institute, Russian Academy of Sciences, 194021 St. Petersburg, Russia}
\affiliation{$^{3}$St. Petersburg National Research University ITMO, 197101 St. Petersburg, Russia.}
\affiliation{$^{4}$Department of Physics and Materials Science, City University of Hong Kong,  Hong Kong S.A.R. }

\date{\today}

\begin{abstract}

The recombination and spin dynamics of excitons in colloidal CdTe
nanocrystals (NCs) are studied by time-resolved photoluminescence in
high magnetic fields up to 15~T and at cryogenic temperatures. The
recombination decay shows a nonexponential temporal behavior, with
the longest component corresponding to the dark excitons having
260~ns decay time at zero magnetic field and 4.2~K temperature. This
long component shortens to 150~ns at 15~T due to the
magnetic-field-induced mixing of the bright and dark exciton states.
The spin dynamics, assessed through the evolution of the
magnetic-field-induced circular polarization degree of the
photoluminescence, has a fast component shorter than 1~ns related to
the bright excitons and a slow component of 5-10~ns associated with
the dark excitons. The latter shortens with increasing magnetic
field, which is characteristic for a phonon-assisted spin relaxation
mechanism. The relatively low saturation level of the associated
magnetic-field-induced circular polarization degree of $-30\%$ is
explained by a model that suggests the CdTe NCs to constitute an
ensemble of prolate and oblate NCs, both having a structural
quantization axis. The exciton $g$-factor of 2.4-2.9 evaluated from
fitting the experimental data in the frame of the suggested approach
is in good agreement with the expected value for the dark excitons
in CdTe NCs.

\end{abstract}
\pacs{73.21.La, 78.47.jd, 78.55.Et, 78.67.Hc}


\maketitle

\section{Introduction}
\label{SecI}

Colloidal
thiol-capped CdTe nanocrystals (NCs) are well known example of the
colloidal quantum dots directly synthesized in aqueous solution
\cite{Tang2002, Gaponik2002, Byrne2007, Rogach2007}. Due to the  great variety
of surface functionalities offered by various thiols, which can be as
capping  molecules at the synthetic stage, a large variety of composite materials employing this kind of nanocrystals as building blocks can be designed. Due to their high photoluminescence quantum yields these NCs have found applications in various fields ranging from light harvesting and energy transfer \cite{Rogach2011} to biotechnology\cite{Zhang2009, Lovric2005, Han2008, Rogach2010}. Colloidal NCs
have also reach potential applications in the devices exploring the
spin degree of freedom.\cite{Spin_book,Vincenzo2000,Glozman2001}

The effect of an applied magnetic field on the exciton recombination
dynamics has been studied for CdSe
\cite{Efros1996,Johnston-Halperin2001,Furis2005,Liu2013}, PbSe
\cite{Schaller2010} and CdTe \cite{Blokland2011} NCs. The impact is
most pronounced at low temperatures, when in equilibrium the dark
exciton states are predominantly populated. The recombination
dynamics show a fast component contributed by recombination of
bright excitons and their fast relaxation to the dark state and a
slow component contributed by dark exciton recombination. The
magnetic field mixes the bright and dark exciton states, resulting
in shortening of the slow decay component and vanishing of the fast
component. This characteristic  behavior in magnetic field looks
similar to the behavior induced by a temperature increase due to
thermal population of the bright exciton state.

Exciton spin relaxation times have been measured for CdSe NCs in
magnetic fields by means of time-resolved photoluminescence. For
neutral NCs the exciton spin dynamics was faster than 1~ns
\cite{Johnston-Halperin2001,Gupta2002, Furis2005,Liu2013}, while in
charged NCs the spin relaxation of the negatively charged exciton
can be as long as 60~ns \cite{Liu2013}. Time-integrated studies of
the magnetic-field-induced circular polarization degree of the
exciton emission from CdSe NCs have been reported~\cite{Wijnen2008}
and it has been also shown that the polarization degrees are similar
for isolated NCs and NC aggregates~\cite{Blumling2011}. In CdTe NCs
the exciton spin dynamics have not been studied so far.

In this manuscript we report on magneto-optical studies of the recombination
and spin dynamics of excitons in an ensemble of colloidal CdTe NCs.


\section{Experimentals}
\label{SecII}

The thiol-capped colloidal CdTe NCs with an average core diameter of
3.4 nm were synthesized in water following the method described in
Ref.~\onlinecite{Rogach2007}. For the optical experiments performed
at cryogenic temperatures, a solution of CdTe NCs was drop-cast on a
glass slice and dried in air. 
The glass slice was mounted in a sample holder and inserted into a
cryostat equipped with a 15 T superconducting magnet. The magnetic
field, \textbf{B}, was applied in the Faraday geometry, i.e. it was
oriented perpendicular to the glass slice and parallel to the light
wave vector. A circular polarizer inserted between the sample and
the detection fiber allowed us to analyze the $\sigma^{+}$ and
$\sigma^{-}$ polarization of the emission by inverting the magnetic
field direction. The sample was in contact with helium gas so that
the bath temperature could be varied from $T=4.2$~K up to 300~K.

Photoluminescence (PL) was excited and collected through multimode
optical fibers. The collected signal was dispersed with a 0.55~m
spectrometer. Time-integrated PL spectra were measured under
continuous-wave (CW) laser excitation with a photon energy of
3.33~eV (wavelength 372~nm) and detected with a
liquid-nitrogen-cooled charge-coupled-device camera. We denote them
as steady-state PL spectra.

For time-resolved measurements the sample was excited by a
picosecond pulsed laser (photon energy 3.06~eV, wavelength 405~nm,
pulse duration 50~ps, repetition frequency 150 to 500~kHz). Here the
PL signal was sent through the spectrometer and detected by an
avalanche photodiode (time response 50~ps) connected to a
conventional time-correlated single-photon counting module. The
instrument response function of the whole setup was limited by the
optical fiber dispersion to a resolution of 800~ps. All measurements
were performed at low excitation densities of about 0.1~mW/cm$^{2}$
to suppress any multiexciton contributions to the emission
spectra.

Analysis of the PL circular polarization degree induced by an
external magnetic field is a powerful tool to investigate the spin
levels of the exciton complexes, to identify their charging state
and to obtain information on the spin dynamics, see, e.g.,
Refs.~\cite{Liu2013, Dunker2012} and references therein. The degree
of circular polarization (DCP) is defined by:
\begin{eqnarray}
P_\text{c}(t) = \frac{I^{+}(t)-I^{-}(t)}{I^{+}(t)+I^{-}(t)}.
\label{EQ:equation1}
\end{eqnarray}
Here $I^{+}(t)$ and $I^{-}(t)$ are the $\sigma^{+}$  and
$\sigma^{-}$ polarized PL intensities, respectively, measured at
time delay $t$ after pulsed excitation~\cite{comment1}. A
saturation of $P_\text{c}(t)$  at times much longer than the exciton
spin relaxation time gives the equilibrium circular polarization
degree $P_\text{c}^{eq}(B)$.

The time-integrated DCP can be evaluated by integrating the
corresponding PL intensities over time:
\begin{eqnarray}
P_\text{c}^\text{int} = \frac{\int dtI^{+}(t)-\int dtI^{-}(t)}{\int dtI^{+}(t)+\int dtI^{-}(t)}.
\label{EQ:equation2}
\end{eqnarray}
In case of CW excitation the measured polarization degree corresponds to $P_\text{c}^\text{int}$.

The magnetic-field-induced DCP is caused by exciton thermalization
among the Zeeman spin levels. While the equilibrium polarization
degree $P_\text{c}^{eq}(B)$ is controlled solely by the thermal
equilibrium population of the Zeeman spin levels, the
time-integrated polarization $P_\text{c}^\text{int}$   depends also
on the ratio of the exciton spin relaxation time $\tau_\text{s}$ and
the exciton lifetime $\tau$:
\begin{eqnarray}
P_\text{c}^\text{int}(B)=
\frac{\tau}{\tau+\tau_\text{s}}P_\text{c}^{eq}(B).
\label{EQ:equation3}
\end{eqnarray}
If $\tau_\text{s} \ll \tau$ the experimentally measured
$P_\text{c}^\text{int}(B)$ coincides with  $P_\text{c}^{eq}(B)$,
while otherwise their difference is controlled by the dynamical factor
$d=\tau /(\tau+\tau_\text{s})$.

The magnetic field dependence of $P_\text{c}^\text{int}(B)$ can be
rather complicated as $\tau_\text{s}$ and $\tau$ can also be
functions of magnetic field. In addition, one has to take into
account the exciton level degeneracy and exciton fine structure in
zero magnetic field, which depends on the crystal symmetry of the
semiconductor material and on the nanocrystal shape. In spherical
hexagonal CdSe NCs or prolate cubic CdTe NCs the hole states are
split by the anisotropy field, and the exciton ground state has
angular momentum projections of $\pm 2$ onto the quantization axis.
The exciton Zeeman splitting depends on the angle between the
quantization axis and the magnetic field direction. The analysis of
DCP for an ensemble of such anisotropic NCs with random orientation
of their axes relative to the magnetic field is presented in
Refs.~\cite{Johnston-Halperin2001,EfrosCh3} for neutral and in
Ref.~\cite{Liu2013} for charged excitons (trions).

Note that in ideally round-shaped NCs with cubic lattice structure
there is no preferential structural quantization axis,
while such an axis becomes well-defined when the NC shape is
deformed to be prolate or oblate and/or for materials with lower
crystal symmetry, e.g., wurtzite CdSe. The shape of thiol-capped NCs
with cubic-phase zinc blend crystal structure had been determined
not to be spherical, but rather tetrahedron or truncated
tetrahedron-like \cite{Shanbhag2006,Tang2006}. While the symmetry of a
tetrahedron is the same as for spherical dots, truncated tetrahedrons
can be considered as prolate or oblate particles.

\section{Results}
\label{SecIII}

\begin{figure}[htc]
\includegraphics[width=\linewidth]{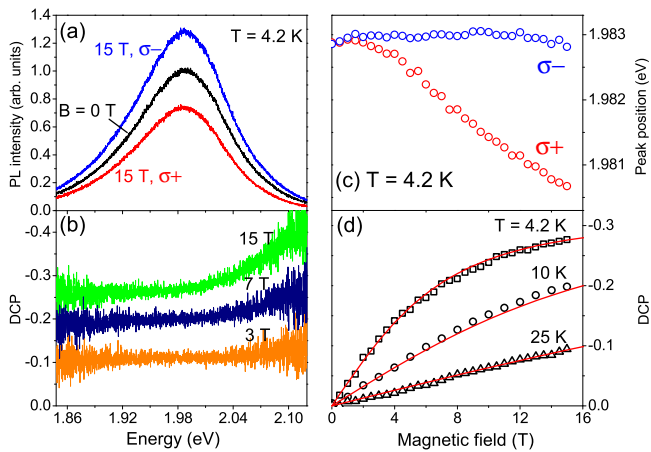}
\caption{(a) PL spectra of 3.4~nm CdTe NCs at $B=0$ (black) and 15~T
(red for $\sigma^+$ polarization and blue for $\sigma^-$). (b)
Spectral dependence of DCP at $B=3$, 7 and 15~T and $T=4.2$~K. (c)
Magnetic field dependence of peak energies of the $\sigma^+$ and
$\sigma^-$-polarized components of the PL spectrum.  (d) Magnetic
field dependence of the spectrally- and temporally-integrated DCP
measured at $T=4.2$, 10 and 25~K. Lines are fits using
Eqs.~\eqref{EQ:mix1} and \eqref{EQ:mix2} with $g_\text{F}=2.4$
and 2.9, respectively. The fits performed along the two equations
coincide with each other to high accuracy and, therefore, are shown
by a line for each temperature data set.} \label{fig:pl}
\end{figure}

Photoluminescence spectra of the CdTe NCs measured under unpolarized
CW excitation are shown in Fig.~\ref{fig:pl}(a). At zero magnetic
field the PL spectrum has its maximum at 1.988~eV with a full width
at half maximum (FWHM) of 120~meV, caused by the NCs size
dispersion.

In an external magnetic field the PL becomes circularly polarized
with an intensity redistribution in favor of the $\sigma^-$
component, i.e. the sign of the circular polarization degree defined
by Eq.~\eqref{EQ:equation2} is negative. The polarization is caused
by exciton thermalization among the Zeeman spin-split levels. At the
PL maximum it reaches a value of $P_\text{c}^\text{int}=-0.28$ at
$B=15$~T. $P_\text{c}^\text{int}$ varies only little across the
emission band as shown in Fig.~\ref{fig:pl}(b), with some small
increase at the high energy tail of the PL spectrum and in
magnetic fields exceeding 6~T.

The magnetic field dependences of the spectrally- and
temporally-integrated DCP measured under CW excitation at three
different temperatures are given in Fig.~\ref{fig:pl}(d). They show
the typical behavior for magnetic-field-induced DCP with a linear
increase in low magnetic fields and saturation in the high magnetic
field limit, as clearly seen for $T=4.2$~K. The decrease of the DCP
at elevated temperatures is also typical, as the occupation of
Zeeman levels is controlled by the ratio of Zeeman splitting to
thermal energy $k_BT$ \cite{OO_book, Liu2013}. As one can see from
Eq.~\eqref{EQ:equation3}, the DCP depends both on the dynamical
factor $\tau /(\tau+\tau_\text{s})$ and on the equilibrium
polarization $P_\text{c}^\text{eq}(B)$. The time-resolved
experiments presented below show that for the studied CdTe NCs
$\tau_\text{s} \ll \tau$ and, therefore, the experimentally measured
$P_c(B)$ corresponds to $P_\text{c}^\text{eq}(B)$.

Figure~\ref{fig:pl}(c) shows the magnetic field dependences of the
PL peak energies of the $\sigma^+$ and $\sigma^-$ polarized spectra.
The energy splitting between them reaches 2.3~meV at $B=15$~T. Here
it is unusual that the $\sigma^-$ component with stronger intensity
is higher in energy. Such a behavior has been observed also for CdSe
NCs \cite{Furis2005, Wijnen2008}, but a plausible explanation is
still missing and further studies are needed to clarify that. 

\begin{figure}[htc]
\includegraphics[width=0.9\linewidth]{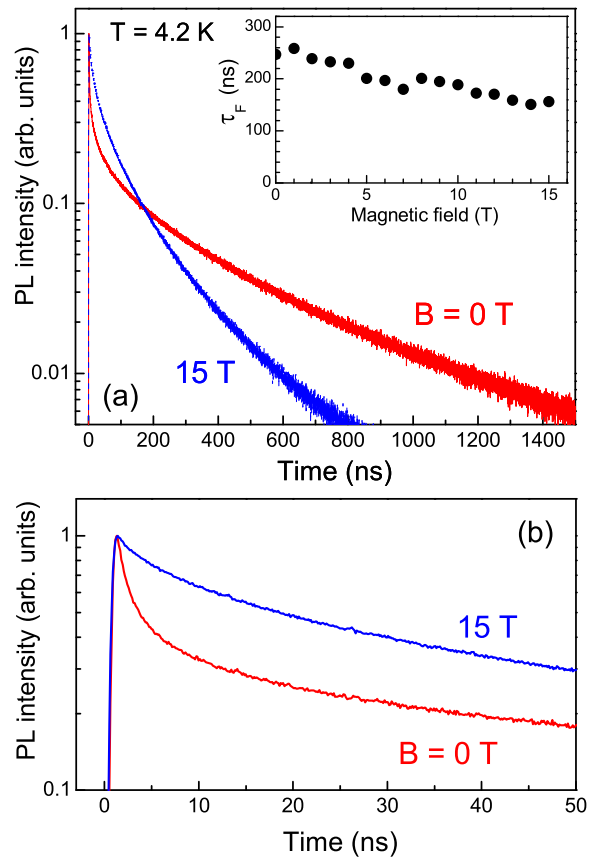}
\caption{(a) Dynamics of spectrally-integrated PL intensity of
3.4~nm CdTe NCs measured at zero and 15~T magnetic field. The
initial dynamics after pulsed excitation are given in more detail in
panel (b). The magnetic field dependence of the recombination time
of the slow component is shown in the insert.} \label{fig:decay}
\end{figure}

Let us first present the experimental data for the exciton
recombination dynamics. In order to avoid the consideration of
effects related to spectral diffusion caused by F\"orster resonant
energy transfer (FRET) among the NCs, we measured the
spectrally-integrated recombination dynamics in the CdTe NCs, see
Fig.~\ref{fig:decay}. It is known, that FRET modifies the spectral
dependences of the emission dynamics \cite{Rogach2009, Wuister2005,
Lunz2010, Blumling2012}. A similar behavior has been also found
for the studied sample and will be reported elsewhere
\cite{Liu2013ET}.

At zero magnetic field and $T=4.2$~K the exciton PL dynamics shown
in red in Fig.~\ref{fig:decay} has a multiexponential decay. The
fast initial component with a decay time of 2~ns corresponds to the
lifetime of the bright (i.e. optically allowed in the
electric-dipole approximation) excitons $\tau_A$. This time is
contributed by radiative recombination and exciton thermalization
from the bright into the dark state. The slow component with a time
of about 260~ns corresponds to the lifetime of the dark (optically
forbidden) excitons $\tau_F$, whose recombination becomes partially
allowed due to a weak mixing of dark and bright exciton states
caused, e.g., by NC imperfections and surface states.  Note, that
the PL intensity of the slow component appears to be smaller than
that of the fast component, but integrated over time it actually
provides the dominant contribution to the overall emission.

With increasing magnetic field up to 15~T the slow decay component
continuously shortens from 260 down to 150~ns, see blue curve in
Fig.~\ref{fig:decay}(a) and insert. The fast component
simultaneously decreases until it vanishes. The PL decay then still
cannot be described by a single exponential decay, but its initial
part shows a decay with 50~ns characteristic time, considerably
longer than the 2~ns at zero magnetic field, see
Fig.~\ref{fig:decay}(b). This behavior is typical for colloidal NCs
and is due to magnetic-field-induced mixing of bright and dark
exciton states \cite{Johnston-Halperin2001,Furis2005,Blokland2011}.
The mixing takes place only in NCs with symmetry further reduced by
the magnetic field, i.e., in NCs where the field is not parallel to the spin
quantization axis. From the strong magnetic field effect on the
recombination dynamics shown in Fig.~\ref{fig:decay}, we conclude
that the dominant fraction of NCs in the ensembles deviates from the
ideal spherical shape so that it has a quantization axis. This
conclusion, as will be discussed below, is supported by the results
of polarized PL experiments.

\begin{figure}[htc]
\includegraphics[width=0.9\linewidth]{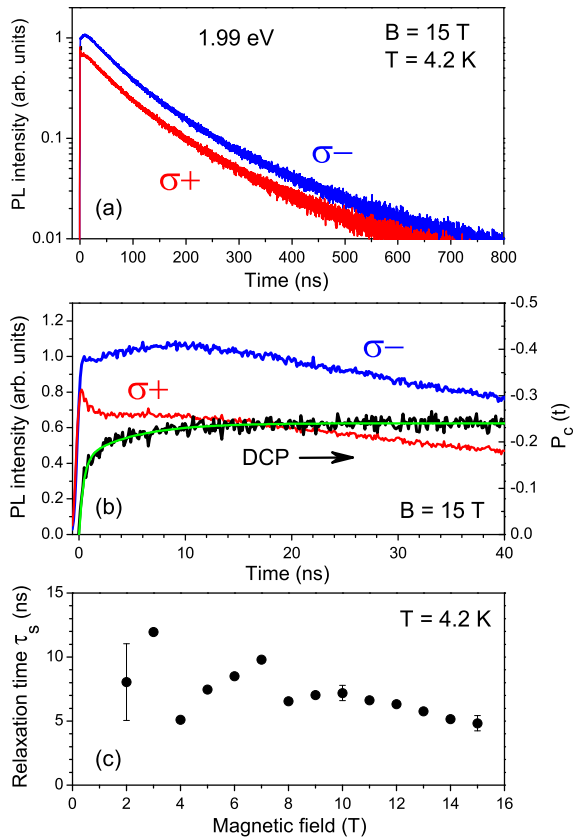}
\caption{Exciton spin dynamics of 3.4~nm CdTe NCs. The two upper
panels give the recombination dynamics of the two circularly
polarized components measured at the PL maximum for $B=15$~T, shown
in (a) on a logarithmic scale for an extended time interval and in
(b) on a linear scale for the initial time interval, along also with
the DCP obtained from these data. (c) Magnetic field dependence of
the exciton spin relaxation time evaluated from the DCP dynamics.
The times for the slower component obtained from bi-exponential fits
of the DCP dynamics are shown. } \label{fig:dcp}
\end{figure}

The experimental data for the exciton spin dynamics are collected in
Fig.~\ref{fig:dcp}. There the recombination dynamics of the two
circularly polarized components measured at the PL band maximum
(1.99~eV) at $B=15$~T are shown in panel (a). One can see, that an
intensity difference between the components is established shortly
after the excitation pulse, after which the two signals decay
roughly parallel to each other. The initial dynamics during 40~ns is
shown in more detail in Fig.~\ref{fig:dcp}(b) together with the
resulting DCP dynamics determined after Eq.~\eqref{EQ:equation1}.
The small rise in PL intensity observed for the $\sigma^-$
polarization during initial 10~ns is due to the energy transfer, see
Ref.~\cite{Liu2013ET} for details. The DCP given by the black line
has a fast initial increase from zero up to about $-0.15$ within
times shorter than 1~ns, which is below the time resolution of the
used experimental setup. It is followed by a slower increase
reaching $-0.25$ at 40~ns and $-0.28$ up to one microsecond. From
the bi-exponential fit shown by the green line in
Fig.~\ref{fig:dcp}(b) we evaluate a slow spin relaxation time of
5~ns at $B=15$~T. The bi-exponential fit was done with the function
\begin{eqnarray}
P_\text{c}(t)= P_\text{c}^\text{eq}(B)\left[1-\text{exp}(-t/\tau_\text{s})\right] + \nonumber  \\
A \left[\text{exp}(-t/\tau_\text{s})-\text{exp}(-t/\tau_\text{fast})\right] \, ,
\label{EQ:trdcpfit}
\end{eqnarray}
where the first term describes the emergence of DCP towards its
equilibrium value $P_\text{c}^\text{eq}$ and the second term
describes its fast initial rise with a characteristic time
$\tau_\text{fast}$  for times shorter than $\tau_{A}$.

The magnetic field dependence of the slow spin relaxation time
$\tau_\text{s}$ is plotted in Fig.~\ref{fig:dcp}(c). It shows a
steady decrease with increasing magnetic fields from about 10~ns
down to 5~ns. Note that in weak magnetic fields the error bars for
the data are larger, as the DCP values become smaller. The accuracy
of the method improves at higher magnetic fields. The spectral
dependence of the DCP rise time is rather weak, e.g., at $B=15$~T it
varies from 4.5~ns at the high energy flank up to 5.5~ns at the low
energy flank of the PL band.

The DCP dynamics consisting of the fast and slow components can be
explained by the spin relaxation contributions of the bright and
dark excitons. The fast component with the spin relaxation time
shorter than 1~ns is provided by the bright exciton relaxation. This
process is ultimately limited by the lifetime of the bright exciton,
which can be as short as 2~ns. Therefore, the slow spin dynamics
with flip times of 5-10~ns can be provided only by spin relaxation
of the dark excitons.  Note that this spin relaxation time is
considerably shorter than the dark exciton lifetime of 150-260~ns.
For this condition ($\tau_\text{s} \ll \tau_\text{F}$) the
experimentally measured DCP under CW excitation $P_\text{c}(B)$ as
used in Fig.~\ref{fig:pl}(d) coincides with the equilibrium spin
polarization $P_\text{c}^{eq}(B)$.

\section{Modeling and Discussion}

Before we turn to the theoretical model, let us briefly summarize
the experimental results. We have shown that in CdTe NCs the
magnetic-field-induced circular polarization the of PL reaches a
saturation level of about $-0.30$ at $T=4.2$~K. The spin relaxation
time of the dark excitons is considerably shorter than its lifetime
($\tau_\text{s} \ll \tau_\text{F}$), so that the polarization under
CW excitation shown in Fig.~\ref{fig:pl}(d) corresponds to the
thermal equilibrium population of the exciton Zeeman levels. The
strong dependence of the recombination dynamics on magnetic field
evidences that most of the NCs have a quantization axis due to
geometry deviation from the spherical shape.

One of the experimental results that we need to explain is the
rather low saturation level of $P_\text{c}=-0.30$ in the limit of
high magnetic fields. Depending on the NC spin structure that is
determined, e.g., by the NC nonspherical shape, different levels of DCP
saturation are expected:

(i) In prolate CdTe NCs, the exciton fine structure is similar to
that of spherical or oblate hexagonal CdSe NCs. The lowest exciton
state is the dark exciton with spin projections on the quantization
axis $\pm 2$, while the states  $\pm1^L$ (here the bright exciton
due to the anisotropy induced admixture of the $\pm1^U$ state) and
$0^L$ are shifted to higher
energies~\cite{EfrosCh3,Efros1996,Blokland2011}.  If the condition
$\tau_\text{s} \ll \tau_\text{F}$ is fulfilled, then the DCP of the
dark exciton is determined only by its Zeeman splitting and the
temperature. The DCP saturation level in high magnetic fields should
be $\pm0.75$ and may become reduced to $\pm0.625$ in case of strong
nonradiative recombination \cite{Johnston-Halperin2001}, here the sign of DCP depends on the exciton $g$-factor. In the
other limiting case $\tau_\text{s} \gg \tau_\text{F}$  or in the
intermediate cases the value of time-integrated DCP will be reduced
by the dynamical factor of the dark exciton
$d_\text{F}=\tau_\text{F}/(\tau_\text{F}+\tau_\text{s})$. However,
the relaxation from bright to dark exciton might transfer the
exciton polarization gained during the relaxation of the bright
exciton to the polarization of the dark exciton and thus contribute to the observed DCP. This situation has not been considered theoretically up to now.

(ii) In oblate CdTe NCs the dark exciton ground state has spin
$0^L$ \cite{EfrosCh3,Efros1996} and, therefore, does not split in
external magnetic field. As long as this state is the lowest one,
i.e. it is not crossed by the next lying bright exciton state with
$\pm 1^L$ that becomes split with increasing magnetic field, the DCP
value is zero.

In the following, we first revisit the case of prolate NCs and show
that even when accounting for the four level system with a transfer
of spin polarization from bright to dark excitons the saturation DCP
level does not significantly reduce. Therefore, none of the
considered scenarios can explain our experimental data. The small
saturation level of $P_\text{c}^{int}$ can, however, be explained by
a model suggesting that the NC ensemble consists of a mixture of
prolate and oblate NCs.

In nonspherical prolate CdTe NCs, the total angular momentum of the
hole is aligned along the quantization axis originating from the
shape anisotropy. Due to the exchange interaction between electron
and hole, the electron spin becomes also pinned along this
quantization axis. As a result the exciton $g$-factors for both dark
and bright excitons are anisotropic and the exciton Zeeman
splittings depend on the angle between the quantization axis and the
magnetic field direction. Since the quantization axis is randomly
oriented, the DCP of an ensemble of NCs has to be integrated over
all angles. The recombination time of the dark exciton, being
controlled by the dark-bright mixing of the exciton states, also
depends on the magnetic field strength and on the NC orientation
relative to the field direction. This may change the relative
contribution of differently oriented NCs to the integral PL
intensity and to the DCP signal in case of significant nonradiative
recombination. The way to account for such contribution has been
suggested in Ref.~\cite{Johnston-Halperin2001}. We extend this
approach for the case of a four level exciton structure with
spin-dependent relaxation from bright to dark exciton.

Details of the four level model are given in the Appendix. The
time-integrated DCP of an ensemble of randomly oriented, prolate
CdTe NCs can be described by
\begin{eqnarray}
P_\text{c}^\text{int}(B,T)=-\frac{\int^{1}_{0}2x  \rho^\text{int}(B,x,T) \eta_\text{p}(B,x)dx}{\int^{1}_{0}(1+x^{2})\eta_\text{p}(B,x)dx} ,
\label{EQ:DCP}
\end{eqnarray}
where \begin{equation}
\rho^\text{int}=\rho_\text{F}(B,x,T) d_\text{F} + (1/2)  \rho_\text{A}(B,x,T) d_\text{A} (1-d_\text{F}) \, .
\label{rho_int}
\end{equation}
Here $d_\text{A}=\tau_\text{A}/(\tau_\text{A}+\tau_\text{sA})$ is
the dynamical factor of the bright exciton, $\eta_\text{p} = [1 +
\tau_\text{r} / \tau_\text{nr}]^{-1}$ is the quantum efficiency of the
prolate CdTe NCs~\cite{Johnston-Halperin2001}, ($\tau_\text{r}$ and  $\tau_\text{nr}$ are the radiative and nonradiative decay times of the dark exciton in prolate dots).  $x=\cos\Theta$, where $\Theta$ is the angle between the
NC quantization axis and the magnetic field direction.
We have introduced two equilibrium polarizations of bright and dark
exciton populations $\rho_{A(F)}(B,x,T)=\text{tanh}(\Delta
E_{A(F)}/2k_\text{B}T)=\text{tanh}(g_\text{A(F)} \mu_B B
x/2k_\text{B}T)$, controlled by the bright, $g_\text{A}$, and dark,
$g_\text{F}$, exciton $g$-factors. Eq. (\ref{rho_int}) is derived
for the case of equal initial population of bright and dark exciton
states and the extreme case of spin dependent relaxation between
bright and dark excitons (see Appendix).

The  Eq. (\ref{rho_int}) has been obtained for the case $k_\text{B}T
\ll \Delta_\text{AF}$, where $\Delta_\text{AF}$ is the zero field
splitting between the bright and dark excitons (a more general
consideration will be presented elsewhere). Note, that if the bright
and dark exciton $g$-factors have different signs the contributions
from the bright and dark exciton polarizations may partly compensate
each other. However, as we have already established the relatively
fast spin relaxation for the dark exciton in experiment, its
dynamical factor $d_\text{F} \approx 1$ and, therefore, no
contribution from the bright exciton to the DCP is expected.

Thus to explain the experimentally observed low saturation level of
DCP at high magnetic fields we suggest that there is a subset of
oblate NCs in the NC ensemble. It is important to note, that for such
NCs the nonzero recombination rate of the dark $0^L$ exciton state
in zero magnetic field is due to the admixture of bright exciton
states, either $1^{L,U}$ or $0^U$. The rate can be
accelerated by a magnetic field component perpendicular to the
anisotropy axis due to the field-induced admixture of the $1^{L}$
states \cite{EfrosCh3}. (We assume that the Zeeman energy is much
smaller than the energy separation to the upper bright exciton
states.) Thus no strong difference between the magnetic field
dependences of the recombination dynamics is expected for prolate
and oblate NCs and the presence of both kinds of NCs leads to
additional independent contributions to the multi-exponent emission
decay only.

Since the exciton ground state of oblate CdTe NCs does not split in
magnetic field, such NCs contribute only to the total PL intensity
and give no DCP signal. Then, the time-integrated DCP
$P_\text{c}^\text{int}(B)$ of an ensemble consisting of CdTe NCs
with oblate and prolate shapes can be written as:
\begin{eqnarray}
{P^\text{int}_\text{c}}(B,T)=-{\frac{\int^{1}_{0}2x  \rho_\text{F}(B,x,T)  \eta_\text{p}(B,x) dx}{\int^{1}_{0} (1+x^{2}) \left[ \eta_\text{p}(B,x) +q\eta_\text{o}(B,x) \right] dx}},
\label{EQ:mix}
\end{eqnarray}
where $q$ is the ratio of the number of oblate to prolate NCs, and
$\eta_\text{o} = [1 + \tau^\text{o}_\text{r} / \tau_\text{nr}^\text{o}]^{-1}$
is the quantum efficiency of oblate NCs ($\tau^\text{o}_\text{r}$ and  $\tau^\text{o}_\text{nr}$ are the radiative and
nonradiative decay times of the dark exciton in oblate NCs). 

Next, we discuss two extreme cases and use them for fitting the
experimental data in Fig.~\ref{fig:pl}(d). First, in the limit of
high quantum yield for both prolate and oblate NCs we put
$\eta_\text{p} = \eta_\text{o} \approx 1$. Then, Eq.~\eqref{EQ:mix}
reduces to
\begin{eqnarray}
{P^\text{int}_\text{c}}(B,T)=-{\frac{\int^{1}_{0}2x  \rho_\text{F}(B,x,T) dx}{(1+q)\int^{1}_{0} (1+x^{2}) dx}}.
\label{EQ:mix1}
\end{eqnarray}
The time-integrated DCP measured at the three explored temperatures
can be fitted well using Eq.~\eqref{EQ:mix1} (see red lines in
Fig.~\ref{fig:pl}(d)) with two fitting parameters: $q=1.5$ and
$g_\text{F}=2.4$. It is worth to note that not only the absolute
value but also the sign of the exciton $g$-factor can be obtained
from our experiment, as it is closely connected to the DCP sign.

Second, in the limit of low quantum yield ($\tau, \tau^\text{o} \gg
\tau_\text{nr}$) at zero magnetic field, one can assume the quantum
efficiency of prolate (oblate) dots gained in magnetic field to be
$\eta_\text{p(o)}(1-x^2)B^2$~\cite{Johnston-Halperin2001}. Then,
Eq.~\eqref{EQ:mix} reduces to
\begin{eqnarray}
{P^\text{int}_\text{c}}(B,T)=-{\frac{\int^{1}_{0}2x
\rho_\text{F}(B,x,T) (1-x^2)
dx}{(1+q\eta_\text{o}/\eta_\text{p})\int^{1}_{0} (1+x^{2}) (1-x^2)
dx}} . \label{EQ:mix2}
\end{eqnarray}
Fitting the experimental data in Fig.~\ref{fig:pl}(d) with
Eq.~\eqref{EQ:mix2} gives $g_\text{F}=2.9$, which is close to the
first limiting case of 2.4 and  $q\eta_\text{o}/\eta_\text{p}=1$.

By comparing these two extreme cases, we conclude that the dark
exciton $g$-factor in 3.4~nm-diameter CdTe NCs is in the range from
2.4 to 2.9. Its positive sign corresponds to the observed negative
circular polarization induced by the magnetic field. The obtained
value is in good agreement with theoretical estimations of the dark
exciton $g$-factor of 3.7 according to $g_\text{F}=g_e-3g_h$
\cite{EfrosCh3,Efros1996}, where $g_e$ and $g_h$ are the electron
and hole $g$-factors. In bulk CdTe $g_e=-1.6$, however, due to the
strong quantum confinement in 3.4~nm NCs it becomes positive and we
estimate its value using the theory developed in
Ref.~\cite{Rodina2003} to be $g_e=0.7$. The hole $g$-factor is
estimated using the expression developed for a potential of
spherical symmetry (see Refs.~\cite{Gelmont1973,EfrosCh3,Efros1996})
as $g_h = -1$. All material parameters used for the calculations
were taken from Ref. \cite{Aliev1993}. The fact that we obtain from
our fitting a value of the exciton $g$-factor, that is expected for
the dark exciton, approves our assignment of the slow component in
the DCP dynamics to the dark exciton.

The measured values of the spin relaxation times of 5-10~ns for the
dark excitons in CdTe NCs are longer than the spin relaxation times
reported earlier for CdSe/ZnS NCs \cite{Johnston-Halperin2001,
Furis2005}, where times shorter than one ns were found at $B=12$~T
and for CdSe/CdS NCs with CdS shells thinner than 4~nm
\cite{Liu2013}. Note, that in CdSe/CdS NCs with shells thicker than
5~nm the emission is dominated by negatively charged excitons
(trions) having rather long spin relaxation times up to 60~ns,
measured at $B=1$~T and $T=4.2$~K \cite{Liu2013}.

In small neutral colloidal NCs, the exciton spin relaxation between
the $\pm 1^L$ states is mainly driven by the long-range
electron-hole exchange interaction and by the exciton-phonon
interaction \cite{Scholes2006, Takagahara2000, Ma2012}. For the dark exciton the relaxation between the $\pm 2$ sublevels
may be caused by the same mechanisms considering the admixture of
the $\pm 1^L$ states. This explains the longer spin relaxation
within the dark exciton doublet compared to the bright excitons and
its acceleration in magnetic field.


To summarize, the exciton recombination and spin relaxation dynamics
have been studied experimentally for colloidal CdTe NCs at cryogenic
temperatures and in high magnetic fields. The strong modification of
the recombination dynamics in magnetic field evidences that the
studied NCs have a structural quantization axis being either prolate
or oblate in shape. This conclusion is confirmed by the low
saturation level of the magnetic-field-induced circular polarization
of the exciton emission. The spin dynamics of the dark excitons
happen in the range of 5-10~ns with a considerable shortening with
increasing magnetic field. The exciton $g$-factor of 2.4-2.9
evaluated from fitting the experimental data in the frame of the
suggested approach is in good agreement with the value expected for
the dark excitons in CdTe NCs.

\textbf{Acknowledgment.}  The authors are thankful to Al.~L.~Efros
and L.~Biadala for stimulating discussions. This work was supported
by the EU Seventh Framework Programme (Grant No. 237252,
Spin-Optronics), the Deutsche Forschungsgemeinschaft, the MERCUR and
by the Research Grant Council of Hong Kong S.A.R. (project
No.[T23-713/11]). A.V.R. acknowledges  support from the Russian
Foundation for Basic Research (Grant No. 13-02-00888-a).




\section*{Appendix}
\label{App}

Here we consider a four-level system for the exciton in magnetic
field with populations $N^\pm_\text{A}(t)$ (bright exciton Zeeman
states) and $N^\pm_\text{F}(t)$ (dark exciton Zeeman states). The
rate equations for $N^\pm_\text{A}(t)$  and $N^\pm_\text{F}(t)$ are
\begin{eqnarray}
\frac{dN^{+}_\text{A}}{dt} &=& -{N^+_\text{A}}(\Gamma_\text{A}+\gamma^+_\text{A}+\gamma_0+\gamma_\text{th}^+)+ N_\text{A}^{-}\gamma_\text{A}^{-} +N_\text{F}^{+} \gamma_\text{th}^+ \nonumber \\
 &+& G(t) \, , \\
\frac{dN^{-}_\text{A}}{dt} &=& -{N^{-}_\text{A}}(\Gamma_\text{A}+\gamma^{-}_\text{A}+\gamma_0+\gamma_\text{th}^{-})+ N_\text{A}^{+}\gamma_\text{A}^{+} +N_\text{F}^{-} \gamma_\text{th}^{-} \nonumber \\
 &+& G(t) \, ,
\end{eqnarray}
and
\begin{eqnarray}
\frac{dN^{+}_\text{F}}{dt} &=& -{N^+_\text{F}}(\Gamma_\text{F}+\gamma^+_\text{F}+\gamma_\text{th}^+)+ N_\text{F}^{-}\gamma_\text{F}^{-} +N_\text{A}^{+}(\gamma_0+ \gamma_\text{th}^+) \nonumber \\
 &+& G(t) \, , \\
\frac{dN^{-}_\text{F}}{dt} &=& -{N^{-}_\text{F}}(\Gamma_\text{F}+\gamma^{-}_\text{F}+\gamma_\text{th}^{-})+ N_\text{F}^{+}\gamma_\text{F}^{+} +N_\text{A}^{-}(\gamma_0+ \gamma_\text{th}^{-}) \nonumber \\
 &+& G(t) \, .
\end{eqnarray}
Here $\Gamma_\text{A(F)}$ are the bright and dark exciton
recombination rates (comprising radiative, $\Gamma_\text{A(F)}^\text{r}$, and nonradiative, $\Gamma_\text{A(F)}^\text{nr}$,
recombination), $\gamma_0$ is the relaxation rate from bright
exciton to dark exciton at zero temperature, and
$\gamma_\text{th}^{\pm}$ takes into account thermally-induced mixing
of bright and dark states. These rates are related to the exciton
lifetimes by $1/\tau_\text{F}=\Gamma_\text{F}$ and
$1/\tau_\text{A}=\Gamma_\text{A}+\gamma_0$. We consider only the
extreme cases of spin dependent relaxation in which only the process
between the $N_\text{A}^{+}$ and $N_\text{F}^+$ states (or the
$N_\text{A}^{-}$ and $N_\text{F}^-$ states) are possible, both at
zero temperature and thermally induced at elevated temperature. For
these processes only the flip of the electron spin is required,
which can be provided via interaction with acoustic phonons (in
one-phonon processes) or via interaction with surface localized
spins. The cross-relaxation processes between the $N_\text{A}^{+}$
and $N_\text{F}^{-}$ states (or the $N_\text{A}^{-}$ and
$N_\text{F}^+$ states) require the flip of the heavy-hole spin 3/2
and, therefore, are considerably slower \cite{Liu2013}.  The spin
relaxation times for the bright and dark excitons are
\begin{equation}
\frac{1}{\tau_\text{sA}} = \gamma_\text{A}^{+}+\gamma_\text{A}^{-}
\, , \quad \frac{1}{\tau_\text{s}} =
\gamma_\text{F}^{+}+\gamma_\text{F}^{-} \, .
\end{equation}
Here we will consider only the case of low temperatures $k_{B}T \ll
\Delta_\text{AF}\pm 1/2|(g_\text{A}-g_\text{F})\mu_B B|$, where
$\Delta_\text{AF}$ is the zero field splitting between bright and
dark excitons, and assume that $1/2|(g_\text{A}-g_\text{F})\mu_B B|
< \Delta_\text{AF}$. This allows us to neglect the thermally-induced
mixing between bright and dark exciton states and put
$\gamma_\text{th}^{\pm}=0$, while thermalization between the Zeeman
levels of the bright (dark) excitons is taken into account. A more
general consideration will be published elsewhere.

The excitation term $G(t)$ describes the unpolarized equal intensity
pumping of bright and dark exciton states. Such pumping corresponds
to our experimental conditions: nonresonant excitation well above
the exciton energy and fast relaxation of hot excitons to the ground
state, assisted by optical phonons and interaction with the surface
\cite{Klimov2000}. For short pulse excitation one may assume
$G(t)=0$ and complete the model by assuming the initial conditions
$N_\text{A}^+(0)=N_\text{A}^-(0)=N_\text{F}^+(0)=N_\text{F}^-(0)=N_0/4$,
where $N_0$ is the total number of nanocrystals with exciton energy
such that it can be excited by the laser pulse. For CW excitation
$G(t)=G$. We have determined the solutions below for the case of CW
excitation.

To obtain these steady state solutions we take
${dN^{+}_\text{F}}/{dt}={dN^{-}_\text{F}}/{dt}={dN^{+}_\text{A}}/{dt}={dN^{-}_\text{A}}/{dt}=0$
and solve the system of four linear equations for
$N_\text{A}=N_\text{A}^++N_\text{A}^-$,
$N_\text{F}=N_\text{F}^++N_\text{F}^-$,  $\Delta
N_\text{A}=N_\text{A}^+-N_\text{A}^-$ and  $\Delta
N_\text{F}=N_\text{F}^+-N_\text{F}^-$:
\begin{eqnarray}
\frac{N_\text{A}}{N}= \frac{\Gamma_\text{F}}{2\gamma_0+\Gamma_\text{A}+\Gamma_\text{F}} \, , \\
\frac{N_\text{F}}{N}= \frac{\Gamma_\text{A}+2\gamma_0}{2\gamma_0+\Gamma_\text{A}+\Gamma_\text{F}} \, , \\
\frac{\Delta N_\text{A}}{N} \approx \frac{\Gamma_\text{F}}{2\gamma_0}\rho_\text{A}d_\text{A} \, , \\
\frac{\Delta N_\text{F}}{N} \approx \left(\rho_\text{F} d_\text{F} +\frac{1-d_\text{A}}{2}\rho_\text{A}\right) \, ,
\end{eqnarray}
where
$N=N_\text{A}+N_\text{F}=G\tau_\text{F}(2-\tau_\text{A}\Gamma_\text{A})+
G\tau_\text{A}$ and we used the relations $\gamma_0 \gg
\Gamma_\text{A} \gg \Gamma_\text{F}$ to approximate the solutions.
One can see, that under these conditions and for $k_{B}T \ll
\Delta_\text{AF}$ the population of the bright exciton state is much
smaller than for the dark exciton state: $N_\text{A} \ll
N_\text{F}$ and $\Delta N_\text{A} \ll \Delta N_\text{F}$.

Following the theory of Refs. \cite{Johnston-Halperin2001,EfrosCh3}
we assume that the polarization properties of the $\pm 2$ dark
exciton recombination are determined solely by coupling with the
$\pm 1$ bright excitons. We obtain for the intensities of the
$\sigma^+$ and $\sigma^{-}$ polarized PL:
\begin{eqnarray}
I^{+}+I^{-} =   2(1+x^2) \left( \eta_\text{A} N_\text{A} +  \eta_\text{F} N_\text{F} \right) \, ,  \\
I^{+}-I^{-}  = 4x  \left( \eta_\text{A} \Delta N_\text{A}  +  \eta_\text{F} \Delta N_\text{F} \right) \, ,
\end{eqnarray}
where $x=\cos\Theta$ and $\eta_\text{A(F)}=[1+\Gamma_\text{A(F)}^\text{nr}/\Gamma_\text{A(F)}^\text{r}]^{-1}$ are the quantum efficiencies of
the bright and dark excitons. One can see, that if $\eta_\text{A}\approx
\eta_\text{F}$, the contribution from $N_\text{A}$ and $\Delta
N_\text{A}$ to the DCP can be neglected. We arrive at an expression
for the DCP under CW excitation (which is equivalent to the
time-integrated DCP) given by Eqs.~(\ref{EQ:DCP}) and (\ref{rho_int}),
where the quantum efficiency of the prolate NCs $\eta_\text{p}$
corresponds to $\eta_\text{F}$.

\end{document}